\documentclass[copyright,creativecommons]{eptcs}
\providecommand{\event}{E. Formenti (Ed.): AUTOMATA \& JAC 2012} 
\usepackage[psamsfonts]{amssymb}
\usepackage{epsfig}

\title{Universality of One-Dimensional Reversible and Number-Conserving Cellular Automata}
\author{Kenichi Morita
\institute{Department of Information Engineering, Hiroshima University\\
Higashi-Hiroshima, 739-8527, Japan}
\email{km@hiroshima-u.ac.jp}
}
\def\titlerunning{Reversible and Number-Conserving Cellular Automata}
\def\authorrunning{K. Morita}

\newtheorem{thm}{Theorem}

\newtheorem{lem}{Lemma}
\newtheorem{prop}{Proposition}

\begin{document}
\maketitle

\begin{abstract}
We study one-dimensional reversible and number-conserving 
cellular automata (RNCCA) that have both properties of 
reversibility and number-conservation. 
In the case of 2-neighbor RNCCA, Garc\'{i}a-Ramos proved 
that every RNCCA shows trivial behavior in the sense that 
all the signals in the RNCCA do not interact each other. 
However, if we increase the neighborhood size, we can 
find many complex RNCCAs. 
Here, we show that for any one-dimensional 2-neighbor reversible 
partitioned CA (RPCA) with $s$ states, we can construct  
a 4-neighbor RNCCA with $4s$ states that simulates the former. 
Since it is known that there is a computationally universal 
24-state 2-neighbor RPCA, we obtain a universal 96-state 
4-neighbor RNCCA. 
\end{abstract}

\section{Introduction}

A reversible cellular automaton (RCA), and a number-conserving 
cellular automaton (NCCA) are kinds of abstract spatiotemporal 
models that reflect physical properties of reversibility, and 
conservation (of mass, energy, etc.), respectively.  
A reversible and number-conserving cellular automaton (RNCCA) 
is thus a model that has both these properties.  
Though an RNCCA is a very restricted subclass of a CA, its 
behavior can be complex if we increase the neighborhood size 
and the number of states. 

So far, NCCAs have been extensively studied, and various 
properties and characterizations of them have been given 
\cite{BF02,DFR03,FG03,FS07,HT91,KT08}. 
In \cite{More03} Moreira investigated universality and 
decidability of NCCAs. 
As for RNCCAs, Schranko and de Oliveira \cite{SO10} made an 
experimental study on one-dimensional RNCCAs, and showed that 
an $s$-state $n$-neighbor RNCCA rule can be decomposed into 
$s$-state 2-neighbor RNCCA rules when $s$ and $n$ are very small. 
Garc\'{i}a-Ramos~\cite{Gar12} proved that, in the 2-neighbor case 
(i.e., radius 1/2), every RNCCA is a {\em shift-identity 
product cellular automaton} (SIPCA). 
An SIPCA is an RNCCA composed of ``shift CAs" in which 
all signals are right-moving, and ``identity CAs" in which 
all signals are stationary. 
Hence, in general, an SIPCA has both right-moving and stationary  
signals, but each signal is independent to others. 
Namely, every right-moving signal simply goes through stationary 
signals without affecting them. 
Therefore, all the 2-neighbor RNCCAs show trivial behaviors in 
the sense that the signals do not interact each other. 
On the other hand, Imai, Martin and Saito~\cite{IMS12} showed 
that, in the 3-neighbor case (i.e., radius 1), there are RNCCAs 
in which some signals can interact with others, and thus they 
show some nontrivial behavior.   
However, it is not known whether there exists a computationally 
universal 3-neighbor RNCCA. 

In this paper, we investigate the 4-neighbor case (i.e., radius 3/2), 
and prove there is a computationally universal RNCCA. 
We show that for any given 2-neighbor $s$-state reversible 
partitioned CA (RPCA) we can construct a 4-neighbor $4s$-state RNCCA 
that simulates the former.  
On the other hand, it is known that there is a computationally 
universal 2-neighbor 24-state RPCA \cite{Mor11a}, which can simulate 
any cyclic tag system proposed by Cook~\cite{Coo04}. 
By this, we can obtain a universal 4-neighbor 96-state RNCCA. 

Computational universality in a variant of a one-dimensional 
RNCCA was studied by Morita and Imai~\cite{MI01}, but this CA was not 
in the standard framework of NCCAs, because a partitioned CA (PCA) 
was used as an NCCA model. 
Though a reversible PCA is a subclass of a standard RCA, 
the number-conserving property of a PCA is somehow different 
from an NCCA, since each cell of a PCA has several parts. 
Namely, while each cell of a usual NCCA has a single number, that of 
a number-conserving PCA has a ``tuple" of numbers. 
Therefore, this paper gives the first universality result of an RNCCA 
in the standard framework of an NCCA.

\section{Preliminaries}

A one-dimensional {\em cellular automaton (CA)} is 
a system defined by
\[ A = (\mathbb{Z}, Q, N, f, \#). \]
Here, $\mathbb{Z}$ is the set of all integers where cells 
are placed. 
$Q$ is a non-empty finite set of states of each cell. 
$N$ is a {\em neighborhood}, which is an element of 
$\mathbb{Z}^m\ (m=1,2,\ldots)$. 
Hence $N$ can be written as $N=(n_1, \ldots, n_m)$ where 
$n_i \in \mathbb{Z}\ (i\in \{1, \ldots, m\})$. 
$f: Q^m \rightarrow Q$ is a {\em local function} that 
determines a state transition of each cell depending on 
the states of its $m$ neighboring cells. 
$\# \in Q$ is a {\em quiescent state} that satisfies  
$f(\#, \ldots, \#)=\#$. 
If $f(q_1,\ldots,q_m)=q$ holds for $q_1,\ldots,q_m,q\in Q$, we 
call this relation $f(q_1,\ldots,q_m)=q$ a {\em transition rule} of $A$.
Thus $f$ can be described as a finite set of transition rules. 

A {\em configuration} over $Q$ is a mapping \ 
$\alpha : \mathbb{Z} \rightarrow Q$.  
Let ${\rm Conf}(Q)$ denote the set of all configurations over $Q$, i.e., 
${\rm Conf}(Q)=\{\alpha \,|\, \alpha: \mathbb{Z} \rightarrow Q\}$.
A configuration $\alpha$ is called {\em finite} if the set 
$\{x\,|\,x \in \mathbb{Z} \wedge \alpha(x) \neq \# \}$ is finite. 
Otherwise it is {\em infinite}. 
The set of all finite configurations is denoted by ${\rm Conf}_{\rm fin}(Q)$. 
Applying the local function $f$ to all the cells in $\mathbb{Z}$
simultaneously, we can obtain a {\em global function} $F$ of $A$ 
that determines how a configuration changes to another. 
More precisely, $F: {\rm Conf}(Q) \rightarrow {\rm Conf}(Q)$ is 
defined by the following formula. 
\[
 \forall \alpha \in {\rm Conf}(Q),\ x \in \mathbb{Z}: \ 
 F(\alpha)(x) =  
   f(\alpha(x + n_1), \ldots, \alpha(x + n_m) ) 
\]
It means that the next state of a cell at the position $x$ is 
determined by the present states of $m$ cells at the positions 
$x + n_1, \ldots, x + n_m$ using the local function $f$. 
If $N = (-r,-r+1,\ldots,0,\ldots,r-1,r)$ for some natural number  
$r \in \mathbb{N}$, then $A$ is called a CA of {\em radius} $r$. 
If $N = (-r,-r+1,\ldots,0,\ldots,r-2,r-1)$ for some positive 
integer $r \in \mathbb{N} - \{0\}$, then $A$ is called a CA of 
radius $(2r-1)/2$. 
Hereafter, we call a CA of radius $r$ ($(2r-1)/2$, 
respectively) by a $(2r+1)$-neighbor CA ($2r$-neighbor CA). 

Let $A$ be a CA, and $F$ be its global function. 
$A$ is called a {\em reversible CA} (RCA) 
iff $F$ is an injection, i.e., it satisfies the following condition. 
\[ \forall \alpha_1, \alpha_2 \in {\rm Conf}(Q): \ 
   \alpha_1 \neq \alpha_2 \Rightarrow F(\alpha_1) \neq F(\alpha_2) \]
A more detailed description on the definition of an RCA is found 
e.g.~in \cite{Mor08}. 
\medskip

A number-conserving CA is a one such that each cell's state 
is an integer, and their sum in a configuration is conserved 
throughout the evolving process. 
So far, several definitions and characterizations have been 
given for number-conserving CAs \cite{BF02,DFR03,FG03,HT91}. 
Durand, Formenti and R\'{o}ka~\cite{DFR03} proved that the 
three notions, {\em periodic-number-conserving}, \, 
{\em finite-number-conserving}, \, and\, {\em number-conserving} 
(for infinite configurations), are all equivalent. 
In this paper, we employ the notion of finite-number-conserving 
to define a number-conserving CA. 

Let $A = (\mathbb{Z}, Q, N, f, 0)$ be a CA, 
where $Q=\{0,\ldots,s-1\}\ (s \in \mathbb{N}-\{0\})$, and  
$F$ be its global function. 
The CA $A$ is called {\em finite-number-conserving}, 
if the following condition holds. 
\[ 
\forall \alpha \in {\rm Conf}_{\rm fin}(Q):\ 
 \sum_{x \in \mathbb{Z}} \alpha(x) = \sum_{x \in \mathbb{Z}} {F}(\alpha)(x)
\]
A CA is called a {\em number-conserving cellular automaton} (NCCA), if 
it is finite-number-conserving. 
\medskip

A CA that satisfies both reversibility and finite-number-conserving 
conditions is called a {\em reversible number-conserving CA} (RNCCA). 
\medskip

Next, we give a definition of a partitioned CA, because,  
in the next section, we will show a method of converting 
a reversible partitioned CA into an RNCCA.  
A {\em one-dimensional partitioned cellular automaton} (PCA) is defined by
\[ P = (\mathbb{Z}, (Q_1,\ldots, Q_m), (n_1,\ldots,n_m), f). \]
Here, $Q_i$ ($i=1,\ldots,m$) is a non-empty finite set of states of 
the $i$-th part of each cell, and thus the state set of each cell 
is $Q = Q_1 \times \cdots \times Q_m$.  
The $m$-tuple $(n_1,\ldots,n_m) \in \mathbb{Z}^m$ is a neighborhood, 
and $f: Q \rightarrow Q$ is a local function. 

Let ${\rm pr}_i: Q \rightarrow Q_i$ be the projection function 
such that ${\rm pr}_i(q_1,\ldots,q_m)=q_i$ for all 
$(q_1,\ldots,q_m) \in Q$.
The global function \ 
$F: {\rm Conf}(Q) \rightarrow {\rm Conf}(Q)$ of $P$ is defined 
as the one that satisfies the following formula.
\[
 \forall \alpha \in {\rm Conf}(Q), x \in \mathbb{Z}: 
 F(\alpha)(x) =  
   f({\rm pr}_1(\alpha(x + n_1)), \ldots, {\rm pr}_m(\alpha(x + n_m)) ) 
\]

\begin{figure}[b]
\begin{center}
 \epsfig{file=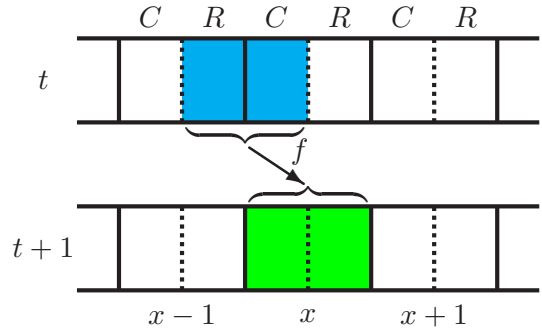, height=43mm}
 \caption{ \label{FIG:pca2space} 
  Cellular space of a one-dimensional 2-neighbor PCA, 
  and its local function $f$.
 }
\end{center}
\end{figure}

By above, one-dimensional PCA of radius 1/2 is defined as follows. 
\[ 
\begin{array}{lll}
P & = & (\mathbb{Z}, (C,R), (0,-1), f) \\
\end{array}
\]
Each cell has two parts, i.e., center and right parts, 
and their state sets are $C$ and $R$. 
The next state of a cell is determined by the present states 
of the center part of this cell, and the right part of the 
left-neighbor cell (not depending on the whole two parts of 
the two cells). 
Figure~\ref{FIG:pca2space} shows its cellular space, and 
how the local function $f$ is applied. 
Note that, here, the neighborhood is $(0,-1)$ 
rather than $(-1,0)$.

It is easy to show the following lemma that states the equivalence of 
local and global injectivity of a PCA \cite{MH89}. 

\begin{lem}
Let $P = (\mathbb{Z}, (Q_1,\ldots, Q_m), (n_1,\ldots,n_m), f)$ 
be a PCA, and $F$ be its global function. 
Then, the local function $f$ is injective, iff the 
global function $F$ is injective. 
\end{lem}
A PCA with an injective local function is thus called a 
{\em reversible PCA} (RPCA).

\section{Converting an RPCA into an RNCCA}

\begin{lem} \label{Lem:Sim_RPCA_by_RNCCA}
For any given one-dimensional 2-neigbor RPCA 
$P = (\mathbb{Z}, (C,R),$ $(0,-1), f )$, 
we can construct a one-dimensional 4-neigbor RNCCA $A$ that 
simulates $P$ and has $4|C|\!\cdot\!|R|$ states. 
\end{lem}

\noindent
{\bf Proof.} \ 
An RNCCA $A$ that simulates $P$ is given as follows.  
\[A = (\mathbb{Z}, \tilde{Q}, (-2,-1,0,1), \tilde{f}, 0 ), \]
where $\tilde{Q} = \{ 0, 1, \ldots, 4|C|\!\cdot\!|R|-1 \}$. 

We need some preparations to define $\tilde{f}$. 
Let $\hat{C}, \check{C}, \tilde{C}, \hat{R}, \check{R}$, 
and $\tilde{R}$ be as follows.  
\[
\begin{array}{lll}
\hat{C}   &=& \{ 2k|R|\ |\ k=0,1,\ldots,|C|-1 \} \\
\check{C} &=& \{ 2(k+|C|)|R|\ |\ k=0,1,\ldots,|C|-1 \} \\
\tilde{C} &=& \hat{C} \cup \check{C} \\
\hat{R}   &=& \{ k\ |\ k=0,1,\ldots,|R|-1 \} \\
\check{R} &=& \{ k+|R|\ |\ k=0,1,\ldots,|R|-1 \} \\
\tilde{R} &=& \hat{R} \cup \check{R} \\
\end{array}
\]
Each element $\tilde{c}\in \tilde{C}$ 
($\tilde{r}\in \tilde{R}$, respectively) is called 
a {\em heavy} ({\em light}) {\em particle}, which is a stationary 
(right-moving) particle in $A$ as explained later. 
The number $\tilde{c}$ ($\tilde{r}$, respectively) itself can be regarded as 
the mass of the heavy (light) particle.  
(Note that readers may think it strange that the particle of mass $0$ 
is both heavy and light. 
Though the mass $0$ could be considered as non-existence of a particle, 
we employ here the above interpretation for simplicity.)
Clearly every element $\tilde{q}\in \tilde{Q}$ is uniquely decomposed into 
a heavy particle and a light particle, and thus the following holds. 
\[
\begin{array}{lll}
\forall \tilde{q}\in \tilde{Q}, 
\exists \tilde{c}\in \tilde{C}, 
\exists \tilde{r}\in \tilde{R}\ 
(\tilde{q} = \tilde{c} + \tilde{r}) \\
\forall \tilde{c}_1,\tilde{c}_2\in \tilde{C}, 
\forall \tilde{r}_1,\tilde{r}_2\in \tilde{R}\  
(\tilde{c}_1 + \tilde{r}_1 = \tilde{c}_2 + \tilde{r}_2\ \Rightarrow\ 
 \tilde{c}_1 = \tilde{c}_2 \wedge \tilde{r}_1 = \tilde{r}_2) \\
\end{array}
\]
We can thus regard each cell of $A$ has a heavy particle and a light particle. 
We define the following functions 
$\tilde{p}_C: \tilde{Q}\rightarrow \tilde{C}$, and 
$\tilde{p}_R: \tilde{Q}\rightarrow \tilde{R}$,  
which give a heavy particle, and a light particle associated 
with a given $\tilde{q}\in \tilde{Q}$. 
\[
\begin{array}{lll}
\forall \tilde{q}\in\tilde{Q}, 
\forall \tilde{c}\in\tilde{C} \ 
(\tilde{p}_C(\tilde{q}) = \tilde{c}\ \Leftrightarrow\ 
 \exists \tilde{r}\in \tilde{R}\,
 (\tilde{q} = \tilde{c} + \tilde{r}) ) \\
\forall \tilde{q}\in\tilde{Q}, 
\forall \tilde{r}\in\tilde{R} \ 
(\tilde{p}_R(\tilde{q}) = \tilde{r}\ \Leftrightarrow\ 
 \exists \tilde{c}\in \tilde{C}\,
 (\tilde{q} = \tilde{c} + \tilde{r}) ) \\
\end{array}
\]
A pair of heavy particles $(\hat{c},\check{c})\in \hat{C}\times\check{C}$ 
(light particles $(\hat{r},\check{r})\in \hat{R}\times\check{R}$, 
respectively) is called a {\em complementary pair}, 
if $\hat{c} + \check{c} = 2(2|C|-1)|R|$\  
($\hat{r} + \check{r} = 2|R|-1$). 
In the following, a complementary pair of heavy (light, respectively) 
particles is used to simulate a state in $C$ ($R$). 
A pair of states $(\tilde{q}_1,\tilde{q}_2) \in \tilde{Q}^2$ is called 
{\em balanced with respect to heavy (light, respectively) particles} 
if $(\tilde{p}_C(\tilde{q}_1),\tilde{p}_C(\tilde{q}_2))$\   
($(\tilde{p}_R(\tilde{q}_1),\tilde{p}_R(\tilde{q}_2))$)  
is a complementary pair. 
The set of all balanced pairs of states $(\tilde{q}_1,\tilde{q}_2)$ with respect 
to heavy (light, respectively) particles is denoted by $B_C$ ($B_R$), i.e.,  
\[
\begin{array}{lll}
B_C = \{(\tilde{q}_1,\tilde{q}_2) \in \tilde{Q}^2 \ |& 
  (\tilde{p}_C(\tilde{q}_1),\tilde{p}_C(\tilde{q}_2))\in \hat{C}\times\check{C} \\
  & \wedge\,\tilde{p}_C(\tilde{q}_1)+\tilde{p}_C(\tilde{q}_2)=2(2|C|-1)|R|\,\}, \\
B_R = \{(\tilde{q}_1,\tilde{q}_2) \in \tilde{Q}^2 \ |&
  (\tilde{p}_R(\tilde{q}_1),\tilde{p}_R(\tilde{q}_2))\in \hat{R}\times\check{R} \\
  & \wedge\,\tilde{p}_R(\tilde{q}_1)+\tilde{p}_R(\tilde{q}_2)=2|R|-1 \}. \\
\end{array}
\]
It is easy to see that, for any $\tilde{\alpha}\in {\rm Conf}(\tilde{Q})$ 
and $x \in \mathbb{Z}$, the following relations hold.
\[
\begin{array}{lll}
(\tilde{\alpha}(x),\tilde{\alpha}(x+1))\in B_C 
\ \Rightarrow \  
(\tilde{\alpha}(x-1),\tilde{\alpha}(x))\not\in B_C \, \wedge\, 
(\tilde{\alpha}(x+1),\tilde{\alpha}(x+2))\not\in B_C \\
(\tilde{\alpha}(x),\tilde{\alpha}(x+1))\in B_R 
\ \Rightarrow \  
(\tilde{\alpha}(x-1),\tilde{\alpha}(x))\not\in B_R \, \wedge \, 
(\tilde{\alpha}(x+1),\tilde{\alpha}(x+2))\not\in B_R 
\end{array}
\]

We now choose bijections 
$\hat{\varphi}_C: C\rightarrow\hat{C}$, and 
$\hat{\varphi}_R: R\rightarrow\hat{R}$ arbitrarily, and fix them hereafter. 
Then define the bijections 
$\check{\varphi}_C: C\rightarrow\check{C}$, \  
$\check{\varphi}_R: R\rightarrow\check{R}$, 
$\hat{\varphi}: C\times R\rightarrow\hat{Q}$, and 
$\check{\varphi}: C\times R\rightarrow\check{Q}$ as follows, where 
$\hat{Q}  =\{ \hat{c}+\hat{r}\ | \ \hat{c}\in\hat{C}, \hat{r}\in\hat{R} \}$, and 
$\check{Q}=\{ \check{c}+\check{r}\ | \ \check{c}\in\check{C}, \check{r}\in\check{R} \}$. 
\[
\begin{array}{lll}
\forall c\in C \ 
(\check{\varphi}_C(c) = 2(2|C|-1)|R|-\hat{\varphi}_C(c) ) \\
\forall r\in R \ 
(\check{\varphi}_R(r) = 2|R|-1-\hat{\varphi}_R(r) ) \\
\forall c\in C, \forall r\in R \ 
(\hat{\varphi}(c,r) = \hat{\varphi}_C(c)+\hat{\varphi}_R(r) ) \\
\forall c\in C, \forall r\in R \ 
(\check{\varphi}(c,r) = \check{\varphi}_C(c)+\check{\varphi}_R(r) ) \\
\end{array}
\]

Now, $\tilde{f}: \tilde{Q}^4 \rightarrow \tilde{Q}$ is defined as follows. 
\begin{eqnarray}
\tilde{f}(\tilde{q}_{-2},\tilde{q}_{-1},\tilde{q}_{0},\tilde{q}_{1})= 
\left\{\!\!
\begin{array}{ll} 
\hat{\varphi}(f(\hat{\varphi}_C^{-1}(\tilde{p}_C(\tilde{q}_{0})),
                \hat{\varphi}_R^{-1}(\tilde{p}_R(\tilde{q}_{-1})))) 
 & \mbox{if} \ (\tilde{q}_{-1},\tilde{q}_{0})\in B_R 
        \wedge (\tilde{q}_{0},\tilde{q}_{1})\in B_C \\
\check{\varphi}(f(\hat{\varphi}_C^{-1}(\tilde{p}_C(\tilde{q}_{-1})),
                  \hat{\varphi}_R^{-1}(\tilde{p}_R(\tilde{q}_{-2}))))
 & \mbox{if} \ (\tilde{q}_{-2},\tilde{q}_{-1})\in B_R 
        \wedge (\tilde{q}_{-1},\tilde{q}_{0})\in B_C \\
\tilde{p}_C(\tilde{q}_{0}) + \tilde{p}_R(\tilde{q}_{-1}) 
 & \mbox{elsewhere}
\end{array} 
\right. \label{Eq:f}
\end{eqnarray}
Let $\tilde{F}$ be the global function induced by $\tilde{f}$. 
For any configuration $\tilde{\alpha}\in {\rm Conf}(\tilde{Q})$, and
for any $y\in\mathbb{Z}$, 
the value $\tilde{F}(\tilde{\alpha})(y)$ is as follows. 
If $(\tilde{\alpha}(y-1),\tilde{\alpha}(y))\in B_R\ \wedge\ 
(\tilde{\alpha}(y),\tilde{\alpha}(y+1))\in B_C$, 
then 
\begin{eqnarray}
\tilde{F}(\tilde{\alpha})(y) &=&
\hat{\varphi}(f(\hat{\varphi}_C^{-1}(\tilde{p}_C(\tilde{\alpha}(y))),
          \hat{\varphi}_R^{-1}(\tilde{p}_R(\tilde{\alpha}(y-1))))), \label{Eq:F1} \\ 
\tilde{F}(\tilde{\alpha})(y+1) &=&
\check{\varphi}(f(\hat{\varphi}_C^{-1}(\tilde{p}_C(\tilde{\alpha}(y))),
          \hat{\varphi}_R^{-1}(\tilde{p}_R(\tilde{\alpha}(y-1))))). \label{Eq:F2} 
\end{eqnarray}
It means the complementary pairs  
$(\tilde{p}_R(\tilde{\alpha}(y-1)),\,\tilde{p}_R(\tilde{\alpha}(y)))$ and  
$(\tilde{p}_C(\tilde{\alpha}(y)),$ $\tilde{p}_C(\tilde{\alpha}(y+1)))$\, 
interact each other, and the state transition of the RPCA $P$ is simulated. 
Thus, the new complementary pair of heavy particles 
$(\tilde{p}_C(\tilde{F}(\tilde{\alpha})(y)),$ 
$\tilde{p}_C(\tilde{F}(\tilde{\alpha})(y+1)))$
is created at the same position as before, while the pair of light particles 
$(\tilde{p}_R(\tilde{F}(\tilde{\alpha})(y)),
\tilde{p}_R(\tilde{F}(\tilde{\alpha})(y+1)))$
appears at the position shifted rightward by one cell.
On the other hand, if 
$\neg((\tilde{\alpha}(y-2),\tilde{\alpha}(y-1))\in B_R \, \wedge\,  
(\tilde{\alpha}(y-1),$ $\tilde{\alpha}(y))\in B_C)) \ \wedge \ 
\neg((\tilde{\alpha}(y-1),\tilde{\alpha}(y))\in B_R \, \wedge\,  
(\tilde{\alpha}(y),\tilde{\alpha}(y+1))\in B_C)) $, 
then 
\begin{eqnarray}
\tilde{F}(\tilde{\alpha})(y) &=&
\tilde{p}_C(\tilde{\alpha}(y)) + \tilde{p}_R(\tilde{\alpha}(y-1)).  \label{Eq:F3}
\end{eqnarray}
The above means the light particle $\tilde{p}_R(\tilde{\alpha}(y-1))$ simply moves rightward 
without interacting with the stationary heavy particle $\tilde{p}_C(\tilde{\alpha}(y))$.
From (\ref{Eq:F1})--(\ref{Eq:F3}), it is easy to see that the following holds 
for all $x\in\mathbb{Z}$. 
\begin{eqnarray}
(\tilde{\alpha}(x),\tilde{\alpha}(x+1))\in B_C 
&\Leftrightarrow&
(\tilde{F}(\tilde{\alpha})(x),\tilde{F}(\tilde{\alpha})(x+1))\in B_C \label{Eq:BC} \\ 
(\tilde{\alpha}(x),\tilde{\alpha}(x+1))\in B_R 
&\Leftrightarrow&
(\tilde{F}(\tilde{\alpha})(x+1),\tilde{F}(\tilde{\alpha})(x+2))\in B_R \label{Eq:BR} 
\end{eqnarray}

\begin{figure}[h]
\begin{center}
 \epsfig{file=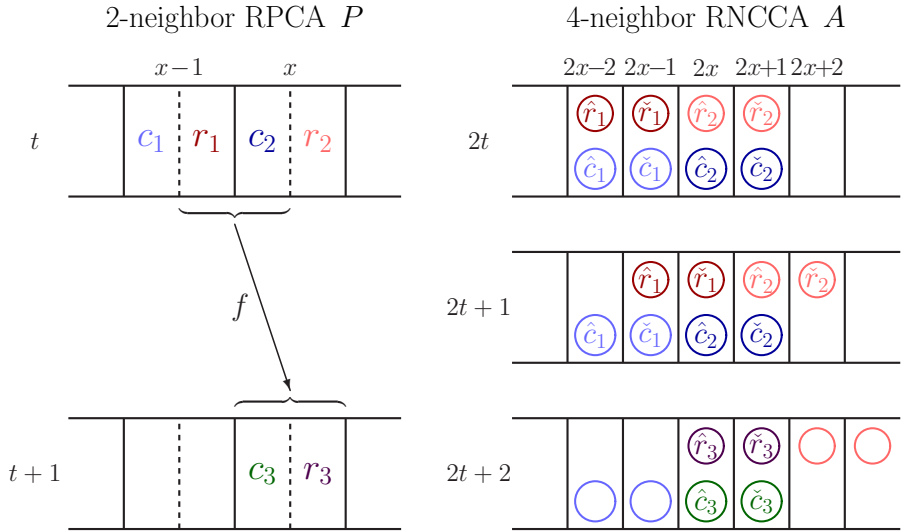, height=70mm}
 \caption{ \label{FIG:RNCCAsim} 
  A simulation process of a 2-neighbor RPCA $P$ by 
  a 4-neighbor RNCCA $A$. 
  The configuration of $A$ at time $2t$ is obtained from 
  that of $P$ at time $t$ by the mapping $\tilde{\tau}$. 
  Here, $(\hat{c}_i, \check{c}_i)$ and $(\hat{r}_i, \check{r}_i)$
  ($i=1,2,3$) are complementary pairs, and thus 
  $\hat{c}_i + \check{c}_i = 2(2|C|-1)|R|$ and 
  $\hat{r}_i + \check{r}_i = 2|R|-1$. 
  }
\end{center}
\end{figure}

First, we show that $A$ can simulate $P$ correctly as described below. 
After that, we will show $A$ is an RNCCA. 
We define a mapping 
$\tilde{\tau}: {\rm Conf}(C\times R) \rightarrow {\rm Conf}(\tilde{Q})$ 
as follows, where $\alpha \in {\rm Conf}(C\times R)$ and $x\in\mathbb{Z}$. 
\begin{eqnarray}
\tilde{\tau}(\alpha)(2x)   &=& \hat{\varphi}(\alpha(x)) \label{Eq:tau1} \\
\tilde{\tau}(\alpha)(2x+1) &=& \check{\varphi}(\alpha(x)) \label{Eq:tau2}
\end{eqnarray}
The configuration $\alpha$ of $P$ is thus represented by 
$\tilde{\tau}(\alpha)$ of $A$ (see Fig.~\ref{FIG:RNCCAsim}). 
We can see   
$(\tilde{p}_C(\tilde{\tau}(\alpha)(2x)),$
 $\tilde{p}_C(\tilde{\tau}(\alpha)(2x+1)))\in B_C$.\ 
However,    
$(\tilde{p}_R(\tilde{\tau}(\alpha)(2x-1)),$ 
$\tilde{p}_R(\tilde{\tau}(\alpha)(2x)))\not\in B_R$ 
for any $x\in\mathbb{Z}$. 
Therefore, by the equation (\ref{Eq:F3}) we have the following. 
\begin{eqnarray*}
\tilde{F}(\tilde{\tau}(\alpha))(2x) &=& 
  \tilde{p}_C(\hat{\varphi}(\alpha(x))) + \tilde{p}_R(\check{\varphi}(\alpha(x-1))) \\
\tilde{F}(\tilde{\tau}(\alpha))(2x+1) &=&
  \tilde{p}_C(\check{\varphi}(\alpha(x))) + \tilde{p}_R(\hat{\varphi}(\alpha(x))) 
\end{eqnarray*}
By above, we can observe  
$(\tilde{p}_C(\tilde{F}(\tilde{\tau}(\alpha))(2x)),\ 
\tilde{p}_C(\tilde{F}(\tilde{\tau}(\alpha))(2x+1)))\in B_C$ and    
$(\tilde{p}_R(\tilde{F}(\tilde{\tau}(\alpha))(2x-1)),$   
$\tilde{p}_R(\tilde{F}(\tilde{\tau}(\alpha))(2x)))\in B_R$, and thus 
the following holds by (\ref{Eq:F1}) and (\ref{Eq:F2}), where
$p_C: C \times R \rightarrow C$ and $p_R: C \times R \rightarrow R$ 
are projection functions, and $F$ is the global function of $P$. 
\begin{eqnarray*}
\tilde{F}^2(\tilde{\tau}(\alpha))(2x) 
 &=& \hat{\varphi}(f(\hat{\varphi}_C^{-1}(\tilde{p}_C(\hat{\varphi}(\alpha(x)))),
     \hat{\varphi}_R^{-1}(\tilde{p}_R(\hat{\varphi}(\alpha(x-1)))))) \\
 &=& \hat{\varphi}(f(p_C(\alpha(x)),p_R(\alpha(x-1)))) \\
 &=& \hat{\varphi}(F(\alpha)(x)) \\
 &=& \tilde{\tau}(F(\alpha))(2x) \\
\tilde{F}^2(\tilde{\tau}(\alpha))(2x+1) 
 &=& \check{\varphi}(f(\hat{\varphi}_C^{-1}(\tilde{p}_C(\hat{\varphi}(\alpha(x)))),
     \hat{\varphi}_R^{-1}(\tilde{p}_R(\hat{\varphi}(\alpha(x-1)))))) \\
 &=& \check{\varphi}(f(p_C(\alpha(x)),p_R(\alpha(x-1)))) \\
 &=& \check{\varphi}(F(\alpha)(x)) \\
 &=& \tilde{\tau}(F(\alpha))(2x+1) 
\end{eqnarray*}
Thus, each evolution step of a configuration of $P$ is correctly simulated 
by $A$ in two steps under the mapping $\tilde{\tau}$. 
Its simulation process is shown in Fig.~\ref{FIG:RNCCAsim}. 

Next, we show that $A$ is an NCCA. 
From the equation (\ref{Eq:f}), we can see mass of a particle is transferred 
within a complementary pair, or simply shifted rightward, or does not change. 
Therefore, it is intuitively obvious that $A$ is an NCCA. 
But, here we show that $A$ has the finite-number-conserving property. 
First, from (\ref{Eq:F1})--(\ref{Eq:F3}), we can derive the following. 
\[
\begin{array}{l}
(\tilde{\alpha}(x),\tilde{\alpha}(x+1))\in B_C \\ \hspace{2em}
\ \Rightarrow \ 
\tilde{p}_C(\tilde{\alpha}(x)) + \tilde{p}_C(\tilde{\alpha}(x+1)) 
= \tilde{p}_C(\tilde{F}(\tilde{\alpha})(x)) + \tilde{p}_C(\tilde{F}(\tilde{\alpha})(x+1)) \\
(\tilde{\alpha}(x-1),\tilde{\alpha}(x))\not\in B_C \wedge 
  (\tilde{\alpha}(x),\tilde{\alpha}(x+1))\not\in B_C \\ \hspace{2em}
\ \Rightarrow \ 
\tilde{p}_C(\tilde{\alpha}(x)) = \tilde{p}_C(\tilde{F}(\tilde{\alpha})(x)) \\
(\tilde{\alpha}(x),\tilde{\alpha}(x+1))\in B_R \\ \hspace{2em}
\ \Rightarrow \ 
\tilde{p}_R(\tilde{\alpha}(x)) + \tilde{p}_R(\tilde{\alpha}(x+1))
= \tilde{p}_R(\tilde{F}(\tilde{\alpha})(x+1)) + \tilde{p}_R(\tilde{F}(\tilde{\alpha})(x+2)) \\
(\tilde{\alpha}(x-1),\tilde{\alpha}(x))\not\in B_R \wedge 
  (\tilde{\alpha}(x),\tilde{\alpha}(x+1))\not\in B_R \\ \hspace{2em}
\ \Rightarrow \ 
\tilde{p}_R(\tilde{\alpha}(x)) = \tilde{p}_R(\tilde{F}(\tilde{\alpha})(x+1)) \\
\end{array}
\]
By above, for each $n\ (=2,3,\ldots)$, there exist $n_0, n_1$ and $n_2$ 
such that $n_i \in \{n, n-1\} \ (i=0,1,2)$, and $n_3 \in \{n-1, n-2\}$ 
that satisfy the following relations.
\begin{eqnarray}
\sum_{x=-n_0}^{n_1} \tilde{p}_C(\tilde{\alpha}(x)) &=& 
  \sum_{x=-n_0}^{n_1} \tilde{p}_C(\tilde{F}(\tilde{\alpha})(x)) \label{Ne:1a} \\ 
\sum_{x=-n_2}^{n_3} \tilde{p}_R(\tilde{\alpha}(x)) &=& 
  \sum_{x=-n_2+1}^{n_3+1} \tilde{p}_R(\tilde{F}(\tilde{\alpha})(x)) \label{Ne:1b} 
\end{eqnarray}

Let $\tilde{\alpha}\in {\rm Conf}_{\rm fin}({\tilde{Q}})$. 
Then, the following equation holds by (\ref{Ne:1a}) and (\ref{Ne:1b}).
\[
 \forall \tilde{\alpha} \in {\rm Conf}(\tilde{Q}): \ 
 \sum_{x \in \mathbb{Z}} \tilde{\alpha}(x) = 
 \sum_{x \in \mathbb{Z}} \tilde{F}(\tilde{\alpha})(x)
\]
Therefore, $A$ is finite-number-conserving, and thus an NCCA. 
\medskip

Finally, we show $A$ is reversible. 
On the contrary we assume it is not. 
Thus, there are two configurations 
$\tilde{\alpha}_1, \tilde{\alpha}_2 \in {\rm Conf}(\tilde{Q})$ 
such that $\tilde{\alpha}_1\neq\tilde{\alpha}_2$ and 
$\tilde{F}(\tilde{\alpha}_1)=\tilde{F}(\tilde{\alpha}_2)$. 
First, we note the following. 
\[
\begin{array}{l}
\forall x \in \mathbb{Z}\ 
( (\tilde{\alpha}_1(x),\tilde{\alpha}_1(x+1))\in B_C \ \Leftrightarrow\ 
  (\tilde{\alpha}_2(x),\tilde{\alpha}_2(x+1))\in B_C )\\
\forall x \in \mathbb{Z}\ 
( (\tilde{\alpha}_1(x),\tilde{\alpha}_1(x+1))\in B_R \ \Leftrightarrow\ 
  (\tilde{\alpha}_2(x),\tilde{\alpha}_2(x+1))\in B_R )\\
\end{array}
\]
If otherwise, $\tilde{F}(\tilde{\alpha}_1)\neq\tilde{F}(\tilde{\alpha}_2)$ 
holds by the relations (\ref{Eq:BC}) and (\ref{Eq:BR}), 
and it contradicts the assumption. 
Since $\tilde{\alpha}_1\neq\tilde{\alpha}_2$, there exists $x_0 \in \mathbb{Z}$ 
such that $\tilde{p}_C(\tilde{\alpha}_1(x_0))\neq\tilde{p}_C(\tilde{\alpha}_2(x_0))$
or $\tilde{p}_R(\tilde{\alpha}_1(x_0))\neq\tilde{p}_R(\tilde{\alpha}_2(x_0))$. 
Here, we prove it only for the case 
$\tilde{p}_C(\tilde{\alpha}_1(x_0))\neq\tilde{p}_C(\tilde{\alpha}_2(x_0))$, 
since the case $\tilde{p}_R(\tilde{\alpha}_1(x_0))\neq\tilde{p}_R(\tilde{\alpha}_2(x_0))$ 
is similarly proved.  
There are three subcases: 
\begin{enumerate}
\item[(i)] $(\tilde{\alpha}_i(x_0-1),\tilde{\alpha}_i(x_0))\in B_R\ \wedge\ 
(\tilde{\alpha}_i(x_0),\tilde{\alpha}_i(x_0+1))\in B_C\ (i=1,2)$, 
\item[(ii)] $(\tilde{\alpha}_i(x_0-2),\tilde{\alpha}_i(x_0-1))\in B_R\ \wedge\ 
(\tilde{\alpha}_i(x_0-1),\tilde{\alpha}_i(x_0))\in B_C\ (i=1,2)$, and 
\item[(iii)] Other than the cases (i) and (ii), i.e., 
$\neg ((\tilde{\alpha}_i(x_0-1),\tilde{\alpha}_i(x_0))\in B_R \ \wedge\  
(\tilde{\alpha}_i(x_0),$ $\tilde{\alpha}_i(x_0+1))\in B_C) \ \wedge\  
\neg((\tilde{\alpha}_i(x_0-2),\tilde{\alpha}_i(x_0-1))\in B_R \ \wedge\  
(\tilde{\alpha}_i(x_0-1),\tilde{\alpha}_i(x_0))\in B_C)\ (i=1,2)$.
\end{enumerate}
The case (i): 
By (\ref{Eq:F1}), the following relations hold.
\begin{eqnarray*}
\tilde{F}(\tilde{\alpha}_1)(x_0) &=&
\hat{\varphi}(f(\hat{\varphi}_C^{-1}(\tilde{p}_C(\tilde{\alpha}_1(x_0))),
          \hat{\varphi}_R^{-1}(\tilde{p}_R(\tilde{\alpha}_1(x_0-1)))))\\
\tilde{F}(\tilde{\alpha}_2)(x_0) &=&
\hat{\varphi}(f(\hat{\varphi}_C^{-1}(\tilde{p}_C(\tilde{\alpha}_2(x_0))),
          \hat{\varphi}_R^{-1}(\tilde{p}_R(\tilde{\alpha}_2(x_0-1)))))
\end{eqnarray*}
From the facts $\tilde{p}_C(\tilde{\alpha}_1(x_0))\neq
\tilde{p}_C(\tilde{\alpha}_2(x_0))$, 
$\varphi, \varphi_C$ and $\varphi_R$ are bijections,  
and $f$ is an injection (because $P$ is a reversible PCA), 
$\tilde{F}(\tilde{\alpha}_1)(x_0) \neq \tilde{F}(\tilde{\alpha}_2)(x_0)$ follows. 
This contradicts the assumption. 
The case (ii): Since it is similar to the case (i), we omit the proof. 
The case (iii): 
By (\ref{Eq:F3}), the following relations hold.
\begin{eqnarray*}
\tilde{F}(\tilde{\alpha}_1)(x_0) &=&
 \tilde{p}_C(\tilde{\alpha}_1(x_0)) + \tilde{p}_R(\tilde{\alpha}_1(x_0-1)) \\
\tilde{F}(\tilde{\alpha}_2)(x_0) &=&
 \tilde{p}_C(\tilde{\alpha}_2(x_0)) + \tilde{p}_R(\tilde{\alpha}_2(x_0-1))
\end{eqnarray*}
Again $\tilde{F}(\tilde{\alpha}_1)(x_0) \neq \tilde{F}(\tilde{\alpha}_2)(x_0)$, 
because $\tilde{p}_C(\tilde{\alpha}_1(x_0))\neq\tilde{p}_C(\tilde{\alpha}_2(x_0))$,  
and this contradicts the assumption. 
By above, we can conclude that $A$ is a reversible NCCA. 
This completes the proof. 
\hfill $\Box$
\bigskip

It has been shown that there is a universal one-dimensional 2-neighbor 
24-state RPCA \cite{Mor11a}. 
This RPCA can simulate 
Any cyclic tag system proposed by Cook~\cite{Coo04} can be simulated 
by this RPCA with infinite but ultimately-periodic configurations.

\begin{prop} {\rm \cite{Mor11a}}\ \label{Prop:URPCA}
There is a computationally universal one-dimensional 2-neighbor 24-state RPCA. 
\end{prop}
From Lemma~\ref{Lem:Sim_RPCA_by_RNCCA} and Proposition~\ref{Prop:URPCA}, 
the next theorem is derived. 
\begin{thm} \label{Thm:URNCCA}
There is a computationally universal one-dimensional 4-neighbor 96-state RNCCA. 
\end{thm}

In \cite{Mor11a}, it is shown that there is a 2-neighbor RPCA 
that directly simulates a given reversible Turing machine. 
Therefore, we can also construct a 4-neighbor RNCCA that directly 
simulates a reversible Turing machine. 
In this case, the RNCCA has ultimately periodic infinite configurations, 
though the configuration of the simulated Turing machine is finite.

\section{Concluding remarks}

In this paper, we proved that any given 2-neighbor RPCA $P$ can be 
simulated by a 4-neighbor RNCCA $A$.  
Thus computation-universality of a 4-neighbor RNCCA is concluded 
in spite of the strong constraints of reversibility and the 
number-conserving property. 
When $A$ simulates $P$, a configuration $\alpha$ of $P$ is kept by 
$\tilde{\tau}(\alpha)$ of $A$ as shown in Fig.~\ref{FIG:RNCCAsim}. 
But, there is no need to define $\tilde{\tau}$ as given in 
the equations (\ref{Eq:tau1}) and (\ref{Eq:tau2}). 
The simulation works well if we use, e.g., the following 
$\tilde{\tau}'$, where each two-cell-block containing 
$\hat{\varphi}(\alpha(x))$ and $\check{\varphi}(\alpha(x))$ 
is separated from the next block by $(k-2)$ $0$-state cells 
($k=3,4,\ldots$). 
\begin{eqnarray}
\tilde{\tau}'(\alpha)(kx)   &=& \hat{\varphi}(\alpha(x)) \nonumber \\
\tilde{\tau}'(\alpha)(kx+1) &=& \check{\varphi}(\alpha(x)) \nonumber \\
\tilde{\tau}'(\alpha)(kx+i) &=& 0 \hspace{3em} (i=2,3,\ldots,k-1) \nonumber
\end{eqnarray}
Furthermore, we can see that, even if the spacing between blocks 
(by $0$-state cells) is non-uniform, the simulation process 
goes correctly (though state transition timing of the cells are 
also non-uniform). 

On the other hand, it is an open problem whether a stronger result 
holds, i.e., whether there is a universal 3-neighbor (radius 1) RNCCA. 
It is also left for the future study to construct an intrinsically 
universal RNCCA. 

\bigskip

\noindent
{\bf Acknowledgement.}\ 
This work was supported in part by JSPS Grant-in-Aid for Scientific Research (C) 
No.~21500015 and No.~24500017. 

%
%
\bibliographystyle{eptcs}
\bibliography{morita}

\end{document}